\documentclass[submission, Phys]{SciPost}
\usepackage{amsmath,braket,mathdots,mathtools,amssymb,xcolor,calligra,color}
\usepackage{graphicx}
\usepackage{hyperref}
\usepackage{lipsum}
\usepackage{bbm}

\usepackage{dsfont}
\usepackage{comment}
\usepackage{latexsym}
\usepackage{times}
\usepackage[normalem]{ulem}

\newcommand{\be}{\begin{equation}}
\newcommand{\ee}{\end{equation}}
\newcommand{\bea}{\begin{eqnarray}}
\newcommand{\eea}{\end{eqnarray}}
\newcommand{\up}{\uparrow}
\newcommand{\down}{\downarrow}

\def\nn{\nonumber\\}

\def\fr#1{(\ref{#1})}
\def\sgn{{\rm sgn}}
\def\eps{\epsilon}

\def\at{\widetilde{\alpha}}
\def\ab{\bar{\alpha}}
\def\bt{\widetilde{\beta}}
\def\bb{\bar{\beta}}

\def\et{\widetilde{e}}

\def\al{\alpha}
\def\boa{\boldsymbol{\alpha}}
\def\bob{\boldsymbol{\beta}}
\def\boc{\boldsymbol{\gamma}}
\def\bod{\boldsymbol{\delta}}
\def\ontop#1#2{\genfrac{}{}{0pt}{}{#1}{#2}}

\makeatletter
\newsavebox{\@brx}
\newcommand{\llangle}[1][]{\savebox{\@brx}{\(\m@th{#1\langle}\)}%
  \mathopen{\copy\@brx\kern-0.5\wd\@brx\usebox{\@brx}}}
\newcommand{\rrangle}[1][]{\savebox{\@brx}{\(\m@th{#1\rangle}\)}%
  \mathclose{\copy\@brx\kern-0.5\wd\@brx\usebox{\@brx}}}
\makeatother
\newcommand{\overbar}[1]{\mkern 1.5mu\overline{\mkern-1.5mu#1\mkern-1.5mu}\mkern 1.5mu}

\begin{document}

\begin{center}
{\Large\bf Yang-Baxter integrable Lindblad equations}
\end{center}
\begin{center}
Aleksandra A. Ziolkowska and Fabian H.L. Essler
\end{center}
\begin{center}
The Rudolf Peierls Centre for Theoretical Physics, Oxford
University, Oxford OX1 3PU, UK
\end{center}
\date{\today}

\section*{Abstract}
{\bf We consider Lindblad equations for one dimensional fermionic models
and quantum spin chains. By employing a (graded) super-operator
formalism we identify a number of Lindblad equations than can be
mapped onto non-Hermitian interacting Yang-Baxter integrable models. Employing Bethe
Ansatz techniques we show that the late-time dynamics of some of these
models is diffusive.
}

\vspace{10pt}
\noindent\rule{\textwidth}{1pt}
\tableofcontents\thispagestyle{fancy}
\noindent\rule{\textwidth}{1pt}
\vspace{10pt}

\section{Introduction}
\label{sec:intro}
Weak couplings to an environment can have very interesting effects on
the dynamics of many-particle quantum systems. In particular they can
result in desirable non-equilibrium steady states
\cite{Diehl,Frank,Weimer,Diehl2,Mazza}. In order to 
arrive at a tractable theoretical description it is customary to
employ a Markovian approximation that assumes that the characteristic
times scales associated with the environment are much shorter
than those of the many-particle system of interest. The absence of a
back action of the system onto its environment then facilitates
a well defined mathematical description of open many-particle
systems. In the quantum case this a priori results in a Markovian
quantum stochastic many-particle system \cite{Jin,Jin2,Jin3,Tilloy},
which is however difficult to analyze. The customary approach in
therefore to focus on the dynamics averaged over the environment, which
leads to a description by the Lindblad master equation \cite{BP} for
the time-dependent reduced density matrix $\rho(t)$ 
\be
\frac{d\rho}{dt}=i[\rho,H]+\sum_a \gamma_a\left[L_a\rho
  L^\dagger_a-\frac{1}{2}\{L^\dagger_aL_a,\rho\}\right]. 
\label{Lindblad}
\ee
Here $H$ is the system Hamiltonian, $L_a$ are jump operators that
encode the coupling to the environment and $\gamma_a>0$. While much
progress has been made in analyzing Lindblad equations for
many-particle systems by employing e.g. perturbative
\cite{perturbation,DiehlKeldysh} and matrix product states methods 
\cite{MPS,MPS2,MPS3,MPS4} it clearly is highly desirable to have exact
solutions in specific, and hopefully representative, cases. In the
context of master equations for classical stochastic many particle systems
an example of such a solvable paradigm is the asymmetric simple
exclusion process \cite{GwaSpohn,Kim,Mallick,dGE,dGE2,Crampe}. In the
quantum case it has been known for some time that certain Lindblad
equations describing many-particle systems can be represented by
Liouvillians that are quadratic in fermionic or bosonic creation and
annihilation operators, which makes it possible to solve them exactly
by elementary means \cite{Prosen2008,Eisert,Cirac,Poletti}. Very
recently examples of Lindblad equations with Liouvillians related
to \emph{interacting} Yang-Baxter integrable models have been found
\cite{Medvedyeva,Lamacraft,Shibata}. This opens the door for bringing
quantum integrability methods to bear on obtaining exact results for
the dynamics of open many-particle quantum systems. An obvious
question is whether the known cases are exceptional, or whether
there are other examples of Yang-Baxter integrable Lindblad
equations. In this work we report on the results of a search for
integrable cases among a particular class of Lindblad equations for
translationally invariant many-particle quantum systems.

\section{Lindblad equations for lattice models}
\label{sec:LE}
We now turn to the precise definition of the class of quantum master
equations we will be interested in. We consider one dimensional
lattice models with local Hilbert spaces that can include bosonic as
well as fermionic degrees of freedom. A basis of the local
Hilbert space is formed by $N$ bosonic and $M$ fermionic quantum
states
\be
|\alpha\rangle_j\ ,\quad \alpha=1,\dots,N+M.
\ee
We denote the fermion parity of the state $|\alpha\rangle_j$ by $\epsilon_\alpha$
\be\label{eq:parity_index}
\epsilon_\alpha=\begin{cases}
0 & \text{if $\alpha$ is bosonic}\\
1 & \text{if $\alpha$ is fermionic}
\end{cases}.
\ee
An orthonormal basis of the full Hilbert space ${\cal H}_L$ on an $L$-site chain is
then given by the states 
\be
|\boldsymbol{\alpha}\rangle\equiv
\otimes_{j=1}^L|\alpha_j\rangle_j\ ,\quad\alpha_j\in\{1,\dots,N+M\}.
\label{basisstates0}
\ee
We define the fermion parity of the states \fr{basisstates0} by
\be
\epsilon_{\boldsymbol{\alpha}}=\sum_{j=1}^L\eps_{\alpha_j}\ .
\label{stateparity}
\ee
A basis of the space of linear operators acting on site $j$ is then
provided by 
\be
E_j^{\alpha\beta}=|\alpha\rangle_j\ {}_j\langle\beta|,\quad
\alpha,\beta\in\{1,\dots,N+M\}.
\label{Eab}
\ee
These are often referred to as Hubbard operators. Their fermion parity
is $\epsilon_\alpha+\epsilon_\beta\text{ mod }2$, i.e. they are fermionic
if either the state $|\alpha\rangle$ or the state $|\beta\rangle$ is
fermionic. The operators $E_n^{\alpha\beta}$ act on the states
$|\boldsymbol{\alpha}\rangle$ as
\bea
E_n^{\alpha\beta}|\boldsymbol{\alpha}\rangle=
(-1)^{(\epsilon_\alpha+\epsilon_\beta)\sum_{j=1}^{n-1}\epsilon_{\alpha_j}}\delta_{\beta,\alpha_n}
|\boldsymbol{\alpha}'\rangle\ ,\quad
\boldsymbol{\alpha}'=\alpha_1,\dots,\alpha_{n-1},\alpha,\alpha_{n+1},\dots,\alpha_L.
\eea
Minus signs are acquired when moving fermionic operators past
fermionic states. The operators defined in this way either commute or
anticommute on different sites
\be
E_j^{\alpha\beta}E_k^{\gamma\delta}=(-1)^{(\epsilon_\alpha+\epsilon_\beta)(\epsilon_\gamma+\epsilon_\delta)}
E_k^{\gamma\delta}E_j^{\alpha\beta}\ ,\quad k\neq j.
\ee
For later convenience we define a \emph{graded permutation operator}
on sites $j$ and $j+1$ 
\be
\Pi_{j,j+1}=\sum_{\alpha,\beta}(-1)^{\epsilon_{\beta}}
E_j^{\alpha\beta}E_{j+1}^{\beta\alpha}\ .
\label{GPO}
\ee
It acts on states as
\bea
\Pi_{j,j+1}|\beta\rangle_j |\alpha\rangle_{j+1}
&=&(-1)^{\epsilon_\alpha\epsilon_\beta}|\alpha\rangle_j |\beta\rangle_{j+1},
\eea
i.e. it permutes the states and generates a minus sign if both states
are fermionic.
\subsection{\texorpdfstring{A useful decomposition for
    $N=n_B^2+n_F^2$, $M=2n_Bn_F$ for integer $n_B$, $n_F$}{Lg}}
\label{ssec:decomp}
In these cases it is possible to decompose the local Hilbert space as
a graded tensor product of two $n=n_B+n_F$-dimensional spaces
\be
|\alpha\rangle=|\at\rangle\otimes |\ab\rangle\ ,\quad
\alpha=1,\dots,N+M\ ,
\ee
where $1\leq \bar{\alpha},\widetilde{\alpha}\leq n$ are expressed in
terms of $\alpha$ by
\be
\ab=\alpha\text{ mod }n+n\delta_{\alpha\text{ mod } n,0}\ ,\qquad
\at={\left \lfloor{\frac{\alpha}{n+1}}\right \rfloor}+1.
\label{abat}
\ee
We note that $\alpha=n(\at-1)+\ab$ and that the fermion parities are
related by $\epsilon_\alpha=\epsilon_{\at}+\epsilon_{\ab}$. Defining operators
\be
\et^{\at\bt}_j=|\at\rangle_j\ {}_j\langle\bt|\ ,\quad
e^{\ab\bb}_j=|\ab\rangle_j\ {}_j\langle\bb|,
\ee
we may express $E_j^{\alpha\beta}$ in the form
\bea
E^{\alpha\beta}_j=|\alpha\rangle\langle\beta|=|\at\rangle|\ab\rangle\langle\bb|\langle\bt|
=(-1)^{\epsilon_{\bt}(\epsilon_{\ab}+\epsilon_{\bb})}\ \et_j^{\at\bt}\ e_j^{\ab\bb}.
\eea
We will use this decomposition in several models considered below.

\subsection{Super-operator formalism for Lindblad equations}
We now consider a Lindblad equation \fr{Lindblad} with a Hamiltonian $H$ and jump
operators $L_a$ acting on ${\cal H}_L$ defined above. We are
ultimately interested in cases where the Hamiltonian density and $L_a$
have local expansions in terms of the $E_j^{\alpha\beta}$. To start
with we will assume for simplicity that all jump operators are
bosonic. The cases where some of the jump operators are fermionic will
be discussed later. The reduced density matrix can be expressed in
terms of the basis states defined above as
\be
\rho=\sum_{\boldsymbol{\alpha},\boldsymbol{\beta}}\rho_{\boldsymbol{\alpha},\boldsymbol{\beta}}|\boldsymbol{\alpha}\rangle\langle\boldsymbol{\beta}|\ .
\ee
The matrix elements are related to particular Green's functions of the
operators $E_j^{\alpha\beta}$
\bea
\rho_{\boldsymbol{\alpha},\boldsymbol{\beta}}=
(-1)^{\sum_{j=1}^{L-1}\sum_{k=j+1}^L\epsilon_{\beta_j}\left(\epsilon_{\beta_k}+\epsilon_{\alpha_k}\right)}\
{\rm  Tr}\left[\rho\ E_L^{\beta_L\alpha_L }\dots E_1^{\beta_1\alpha_1 }
  \right].
\eea
In terms of components the Lindblad equation reads
\bea
\frac{d}{dt}\rho_{\boldsymbol{\alpha},\boldsymbol{\beta}}&=&
i\sum_{\boc}\rho_{\boa,\boc}H_{\boc,\bob}
-H_{\boa,\boc}\rho_{\boc,\bob}\nn
&+&\sum_a\gamma_a\Big\{\sum_{\boc,\bod}\big(L_a\big)_{\boa,\boc}\rho_{\boc,\bod}
\big(L^\dagger_a\big)_{\bod,\bob} 
-\frac{1}{2}\sum_{\boc}\big(L^\dagger_aL_a\big)_{\boa,\boc}\rho_{\boc,\bob}
+\rho_{\boa,\boc}\big(L^\dagger_aL_a\big)_{\boc,\bob}\Big\},\nn
\label{components}
\eea
where we have introduced the following notations for the matrix
elements of an operator ${\cal O}$
\be
\langle\boa|{\cal O}|\bob\rangle={\cal O}_{\boa,\bob}\ .
\ee
We can view the density matrix as a state in a $(N+M)^{2L}$
dimensional Hilbert space ${\cal H}_S={\cal H}_L\otimes{\cal H}_L$ with basis states
\be
\ket{\boldsymbol{\alpha}}|\boldsymbol{\beta}\rrangle=
|\alpha_1\rangle_1\ \dots|\alpha_{L}\rangle_{L}\ 
|\beta_{1}\rangle\!\rangle_{1}\ \dots|\beta_{L}\rangle\!\rangle_{L}\ .
\label{basisstates}
\ee
In these notations we have
\be
|\rho\rangle=\sum_{\boldsymbol{\alpha},\boldsymbol{\beta}}\rho_{\boldsymbol{\alpha},\boldsymbol{\beta}}|\boldsymbol{\alpha}\rangle|\boldsymbol{\beta}\rrangle\ ,
\ee
and the ``wave-functions''
$\rho_{\boldsymbol{\alpha},\boldsymbol{\beta}}$ correspond to Green's
functions in the original problem. The Lindblad equation \fr{components} can be cast in the form
\be
\frac{d|\rho\rangle}{dt}={\cal L}|\rho\rangle\ ,
\label{LESO}
\ee
where the Liouvillian ${\cal L}$ for bosonic jump operators $L_a$ is given by
\be
{\cal L}=-iH+i\bar{H}+\sum_a\gamma_a\left[
L_a\overbar{L_a^\dagger}-\frac{1}{2}\left(L^\dagger_aL_a+\overbar{L^\dagger_aL_a}\right)
\right]. \label{Liou}
\ee
Here we employ notations such that ${\cal O}={\cal O}\otimes\mathds{1}$ and
have defined related operators
$\overbar{\cal  O}=\mathds{1}\otimes\overbar{\cal   O}$ by
\be
\llangle\boldsymbol{{\gamma}}|\overbar{\cal O}|\boldsymbol{{\beta}}\rrangle
=\langle\boldsymbol{\beta}|{\cal O}|\boldsymbol{\gamma}\rangle .
\label{tilde}
\ee
One can easily check that taking the scalar product of \fr{LESO} with
the state
$\langle\!\langle\boldsymbol{\beta}|\langle{\boldsymbol{\alpha}}|$
precisely reproduces \fr{components}. A convenient basis for expanding
operators $\overbar{\cal {O}}$ is constructed in terms of operators
$\widetilde{E}_{n}^{\alpha\beta}$ defined as
\be
\widetilde{E}_{n}^{\alpha\beta}=\mathds{1}\otimes\big(|\alpha\rangle\!\rangle_n
\ {}_n\langle\!\langle\beta|\big).
\ee
These act on basis states according to
\bea
\widetilde{E}_{n}^{\alpha\beta}
|\boldsymbol{\alpha}\rangle|\boldsymbol{\beta}\rrangle&=&
(-1)^{(\eps_\alpha+\eps_\beta)\eps_{\boa}}|\boa\rangle\
\widetilde{E}_{n}^{\alpha\beta}|\bob\rangle\!\rangle\nn
&=&(-1)^{(\eps_\alpha+\eps_\beta)\eps_{\boa}}
(-1)^{(\epsilon_\alpha+\epsilon_\beta)\sum_{j=1}^{n-1}\epsilon_{\beta_j}}\delta_{\beta,\beta_n}
|\boa\rangle\|\boldsymbol{\beta}'\rangle\!\rangle\ ,
\eea
where
$|\boldsymbol{\beta}'\rrangle=\ket{\beta_1}_1,\dots,\ket{\alpha}_n,\dots,\ket{\beta_L}_L$
and $\eps_{\boa}$ has been defined in \fr{stateparity}.
We note that the operators  $E_{n}^{\alpha\beta}$ act on ${\cal H}_S$
as $E_{n}^{\alpha\beta}\otimes\mathds{1}$.

\subsection{Fermionic jump operators}
\label{app:FJO}
If some of the jump operators are fermionic the super-operator
formalism needs to be modified. Let us denote the fermion parity of
the jump operator $L_a$ by $\epsilon_{L_a}\in\{0,1\}$.
When written in components the Lindblad equation still takes the form
\fr{components}. However, the Liouvillian \fr{Liou} is now replaced by
\bea
{\cal L}&=&-iH+i\bar{H}+\sum_{a}\gamma_a\left[
  (- i)^{\epsilon_{L_a}}L_a\overbar{L_a^\dagger}
  -\frac{1}{2}\left(L^\dagger_aL_a+\overbar{L^\dagger_aL_a}\right)
  \right],
\label{LiouF}
\eea
The state representing the density matrix is also modified and now
takes the form
\be
|\rho\rangle=\sum_{\boa,\bob}\rho_{\boa,\bob}
\left[(-i)^{\epsilon_{\boa}}P_++i^{\epsilon_{\bob}}P_-\right]
|\boa\rangle|\bob\rangle\!\rangle\ ,
\label{rhoF}
\ee
where $P_\pm$ are projection operators onto states with even and odd
fermion parity respectively
\be
P_\pm=\frac{1\pm(-1)^F}{2}\ ,\quad
(-1)^F=\prod_{\ell=1}^L\prod_{\ontop{\alpha=1}{\epsilon_\alpha=1}}^{N+M}
(1-2E_\ell^{\alpha\alpha})(1-2\widetilde{E}_\ell^{\alpha\alpha}).
\ee
We have
\be
(-1)^F|\boldsymbol{\alpha}\rangle|\bob\rangle\!\rangle=
(-1)^{\epsilon_{\boldsymbol{\alpha}}+\epsilon_{\boldsymbol{\beta}}}|\boldsymbol{\alpha}\rangle
|\bob\rangle\!\rangle.
\ee
It is straightforward to check that inserting \fr{LiouF} and \fr{rhoF} into
the equation
\be
\frac{d}{dt}|\rho\rangle={\cal L}|\rho\rangle
\ee
and expanding it in a basis of states precisely recovers \fr{components}.
We stress that in our construction both bosonic and fermionic jump
operators can be accommodated as long as any given jump operator has
a definite fermion parity. 

\section{Lindblad equations as non-Hermitian two-leg ladders}
As we are interested in Liouvillians with local densities we focus on
jump operators where the index $a$ runs either over the sites or the
nearest-neighbour bonds of a one dimensional ring. In this setting
$-iH-\sum_a\frac{\gamma_a}{2}L^\dagger_aL_a$ and
$i\bar{H}-\sum_a\frac{\gamma_a}{2}\overbar{L^\dagger_aL_a}$ 
describe interactions along the two legs of the ladder, while
$\sum_a\gamma_a L_a\bar{L}_a^\dagger$ play the role of interactions
between the two legs. 
\subsection{Single-site jump operators}
In translationally invariant situations the most general bosonic
single-site jump operator can be written in the form
\be
\ell_j={\sum_{\alpha,\beta}} \lambda_{\alpha\beta}E_j^{\alpha\beta}\ ,
\ee
where $\lambda_{\alpha\beta}=0$ unless $(\epsilon_\alpha+\epsilon_\beta)\text{ mod }2=0$.
This generates ``interaction terms'' between the two legs of the form
\be
\ell_j\overbar{\ell^\dagger_j}={\sum_{\alpha\beta}}{\sum_{\gamma\delta}}
\lambda_{\alpha\beta}\ \lambda^*_{\gamma\delta}\
E^{\alpha\beta}_{j}\widetilde{E}^{\gamma\delta}_{j}\ .
\ee
The other jump operator terms in the Liouvillian generate
``generalized magnetic field terms'' acting on the two legs 
\bea
\ell^\dagger_j\ell_j+\overbar{\ell^\dagger_j\ell_j}&=&\sum_{\beta,\gamma}\Lambda_{\beta\gamma}
E_j^{\beta\gamma}+{\Lambda}_{\gamma\beta}
\widetilde{E}_j^{\beta\gamma}\ ,\quad
\eea
where
$\Lambda_{\beta\gamma}=\sum_\alpha\lambda^*_{\alpha\beta}\lambda_{\alpha\gamma}$.

\subsection{Single-bond jump operators}
The most general bosonic jump operator acting on a bond takes the form
\be
\begin{split}
L_j=\sum_{\alpha,\beta}\lambda_{\alpha\beta}E_j^{\alpha\beta}+\lambda'_{\alpha\beta}
E_{j+1}^{\alpha\beta}
+\sum_{\alpha,\beta,\gamma,\delta}\mu_{\alpha\beta\gamma\delta}\ E_j^{\alpha\beta}E_{j+1}^{\gamma\delta}\ .
\end{split}
\ee
This gives rise to quartic, cubic and quadratic ``interaction terms''
in the Liouvillian. The resulting explicit expression is presented
in Appendix \ref{f_const}. The extension to fermionic jump operators is
straightforward.
  
\subsection{General form of the Liouvillian}
\label{ssec:genform}
In the following we will consider Liouvillians of the form
\be
{\cal L}=-iH+i\bar{H}+\sum_{j=1}^L\sum_{a}\gamma_a\left[
L^{(a)}_j\overbar{(L^{(a)}_j)^\dagger}-\frac{1}{2}\left((L^{(a)}_j)^\dagger
L^{(a)}_j+\overbar{(L^{(a)}_j)^\dagger L^{(a)}_j}\right)\right],
\label{Lgen}
\ee
where $L^{(a)}_j$ are jump operators that act either on site $j$ or
the bond $(j,j+1)$ and $\gamma_a>0$. Our aim is to identify cases
which are Yang-Baxter integrable. In practice this means that we need
to check whether any of the large number of integrable Hamiltonians
that can be interpreted as two-leg ladder models can be cast in the
particular form \fr{Lgen}. An added complication is that we should
allow for general similarity transformations, i.e. consider
\be
{\cal L}'=S{\cal L}S^{-1}.
\ee
The spatial locality of the Hamiltonian density of the various
integrable models imposes strong restrictions on the possible form of
$S$. Transformations of the form
\be
S=\prod_{j=1}^L S_j\ ,
\ee
where $S_j$ acts non-trivially only on site $j$ are always compatible
with the aforementioned local structure.
\section{Generalized Hubbard models}
The first example of a Lindblad equation that is related to an
interacting Yang-Baxter integrable model was presented in
Ref.~\cite{Medvedyeva}, where it was shown that the Lindblad
equation for a tight-binding chain with dephasing noise can be mapped
onto a fermionic Hubbard model with purely imaginary interactions. We
now briefly review some results obtained in that work. We then show
that the mathematical structure that underlies the integrability of
the Hubbard model quite naturally leads to a connection with a
Lindblad equation.

\subsection{SU(2) Hubbard model}
\label{sec:Hubbard}
The Hubbard Hamiltonian is given by
\be
H=-t\sum_{j=1}^L\sum_{\sigma=\uparrow,\downarrow}c^\dagger_{j,\sigma}c_{j+1,\sigma}+c^\dagger_{j+1,\sigma}c_{j,\sigma}+U\sum_{j=1}^L 
\left[n_{j,\uparrow}-\frac{1}{2}\right]\left[n_{j,\downarrow}-\frac{1}{2}\right],
\ee
where $n_{j,\sigma}=c^\dagger_{j,\sigma}c_{j,\sigma}$. The model is
integrable for any complex value of $U/t$ \cite{book}. In terms of the
notations of section \ref{ssec:decomp} we can choose a basis such that
\be
c^\dagger_{j,\uparrow}=e_j^{21}\ ,\quad
c^\dagger_{j,\downarrow}=\et_j^{21}\ ,\quad
n_{j,\up}=e_j^{22}\ ,\quad
n_{j,\down}=\et_j^{22}\ ,
\label{ccdagger}
\ee
and concomitantly
\bea
H(U)&=&-t\sum_j\left[e_j^{21}e_{j+1}^{12}+\et_j^{21}\et_{j+1}^{12}+{\rm h.c.}\right]
+U\sum_j\left[e_j^{22}-\frac{1}{2}\right]\left[\et_j^{22}-\frac{1}{2}\right]\ .
\eea
\subsubsection{Associated Lindblad equation}
Let us consider a tight-binding model
\be
H_0=-t\sum_j e_j^{12}e_{j+1}^{21}+{\rm h.c.}\ ,
\ee
coupled to an environment by jump operators
\be
L_j=e_j^{22}\ .
\ee
In the super-operator formalism the corresponding Liouvillian
\fr{Liou} is
\bea
{\cal L}(\gamma)=it\sum_j
[e_j^{12}e_{j+1}^{21}-\et_j^{12}\et_{j+1}^{21}+{\rm h.c.}]
+\sum_j\gamma\left[e_j^{22}\et_{j}^{22}-\frac{1}{2}(e_j^{22}+\et_j^{22})\right]\ .
\eea
This is related to the Hubbard Hamiltonian by \cite{Medvedyeva}
\be
{\cal L}(\gamma)=-i{\cal U}^\dagger H(i\gamma) {\cal U}-\frac{\gamma L}{4}\ ,\quad
{\cal U}=\prod_{j=1}^{L/2}(\tilde{e}_{2j}^{11}-\tilde{e}_{2j}^{22}).
\label{U}
\ee

\subsection{Integrable structure of generalized Hubbard models and associated
  Lindblad equations}
\label{sec:intHub}
The Hubbard model was embedded into the general framework of the
Quantum Inverse Scattering Method \cite{vladbook} in seminal work
by Shastry \cite{Shastry,Shastry2}. This construction was subsequently
generalized to other classes of integrable models
\cite{Maa3,Marcio2,DFFR,FFR,Marcio}. The construction is based on an
R-matrix $r_{12}(\lambda)$ acting on the tensor product of two graded
linear vector spaces $V\otimes V$ and a conjugation matrix $C$ acting
on $V$ that fulfil the Yang-Baxter relation
\be
r_{12}(\lambda_1-\lambda_2)
r_{13}(\lambda_1-\lambda_3)
r_{23}(\lambda_2-\lambda_3)=
r_{23}(\lambda_2-\lambda_3)
r_{13}(\lambda_1-\lambda_3)
r_{12}(\lambda_1-\lambda_2)\ ,
\ee
as well as the ``decorated'' Yang-Baxter relation
\be
r_{12}(\lambda_1+\lambda_2)C_1
r_{13}(\lambda_1-\lambda_3)
r_{23}(\lambda_2+\lambda_3)=
r_{23}(\lambda_2+\lambda_3)
r_{13}(\lambda_1-\lambda_3)C_1
r_{12}(\lambda_1+\lambda_2)\ .
\ee
In the cases considered below the $r_{12}(\lambda)$ is given by
\be
r_{12}(\lambda)=\left[\cos^2\big(\frac{\lambda}{2}\big)-\sin^2\big(\frac{\lambda}{2}\big)
C_1C_2\right]\Pi_{12}+\frac{\sin(\lambda)}{2}\left[ \mathds{I}\otimes\mathds{I}-C_1C_2\right],
\ee
where $\Pi_{12}$ is a graded permutation operator \fr{GPO} acting on
$V\otimes V$ and
\be
C=2\hat{\pi}-\mathds{1},
\ee
where $\hat{\pi}$ is a projection operator onto a subspace of $V$. The
R-matrix of an integrable generalized Hubbard model is then obtained by gluing
together two copies \cite{DFFR,FFR,book}
\bea
R_{\langle 12\rangle\langle 34\rangle}(\lambda_1,\lambda_2)&=&r_{13}(\lambda_1-\lambda_2)
{r}_{24}(\lambda_1-\lambda_2)\nn
&+&\alpha(\lambda_1,\lambda_2)
r_{13}(\lambda_1+\lambda_2)C_1 {r}_{24}(\lambda_1+\lambda_2){C}_2.
\eea
Here the function $\alpha(\lambda,\mu)$ is given by
\be
\alpha(\lambda,\mu)=\frac{\cos(\lambda-\mu)\sinh\big(h(\lambda)-h(\mu)\big)}
{\cos(\lambda+\mu)\cosh\big(h(\lambda)-h(\mu)\big)},
\ee
where $h(\mu)$ is a solution of the equation
\be
\sinh\big(2h(\lambda)\big)=U\sin(2\lambda).
\ee
The local Hamiltonian density of the integrable ``fundamental spin model'' \cite{vladbook}
corresponding to this R-matrix is 
\bea
H_{\langle 12\rangle\langle 34\rangle}&=&\frac{d}{d\lambda}\bigg|_{\lambda=u_0}\Pi_{13}\Pi_{24}R_{\langle
  12\rangle\langle 34\rangle}(\lambda,u_0)\nn
&=&\Pi_{13}r'_{13}(0)+\Pi_{24}r'_{24}(0)+\alpha'(u_0,u_0)\Pi_{13}r_{13}(2u_0)C_1
\Pi_{24}r_{24}(2u_0)C_2.
\label{Hamil}
\eea
Here we have generalized the construction of \cite{DFFR} by taking the
logarithmic derivative of the transfer matrix at a shifted point $u_0$
following Ref. \cite{SW,jap_Hubbard}. Importantly the structure of the
Hamiltonians \fr{Hamil} is such that they all can be related to
Liouvillians of Lindblad equations. In the following we discuss a
number of examples.

\subsection{USW model}
\label{sec:USW}
As a first application we consider eqn \fr{Hamil} for the case of the
Hubbard model R-matrix \cite{SW}. The Hamiltonian of these models was
first derived by Umeno, Shiroishi and Wadati in \cite{jap_Hubbard} and
is of the form 
\be
H_{\rm USW}(U)=-\sum_j\left[e_j^{21}e_{j+1}^{12}+\et_j^{21}\et_{j+1}^{12}+{\rm
    h.c.}\right]
+\frac{U}{\cosh\left(2h(u_0)\right)}\sum_j B_{j,j+1}\tilde{B}_{j,j+1}
\label{HUSW}
\ee
where
\be
B_{j,j+1}=\big[\cos^2(u_0)\left(e_j^{11}-e_j^{22}\right)
-\sin^2(u_0)\left(e_{j+1}^{11}-e_{j+1}^{22}\right)
+\sin(2u_0)\left(e_{j}^{21}e_{j+1}^{12}-e_{j+1}^{21}e_{j}^{12}\right)\big],
\ee
and $\tilde{B}_{j,j+1}$ is obtained from ${B}_{j,j+1}$ by replacing
$e_n^{\alpha\beta}\rightarrow\tilde{e}_n^{\alpha\beta}$. Here $u_0$ is
a free (complex) parameter and the function $h(u)$ is fixed by the
requirement 
\be
\sinh\left(2h(u_0)\right)=U\sin(2u_0)\ .
\ee
We note that the operators $e_j^{\alpha\beta}$ are related to spinful
fermion creation and annihilation operators by \fr{ccdagger}.
The Hamiltonian \fr{HUSW} is SO(4) symmetric \cite{jap_Hubbard} and in
particular commutes with the total particle number
\be
\hat{N}=\sum_{j=1}^L e_j^{22}+\tilde{e}_j^{22}.
\ee
\subsubsection{Associated Lindblad equation}
The USW model is related to a Lindblad equation with a tight-binding Hamiltonian
\be
H_0=-\sum_j e_j^{12}e_{j+1}^{21}+{\rm h.c.}\ ,
\ee
and jump operators
\bea
L_j&=&B_{j,j+1},
\label{LUSW}
\eea
where the parameter $u_0$ is taken to be purely imaginary.
In the super-operator formalism the corresponding Liouvillian
\fr{Liou} is
\be
{\cal L}(\gamma)=i\sum_j
\big[e_j^{12}e_{j+1}^{21}-\et_j^{12}\et_{j+1}^{21}+{\rm h.c.}\big]
+\gamma\sum_j\left[ B_{j,j+1}\tilde{B}^*_{j,j+1}-\cos^2(2u_0)\right].
\ee
This is related to the USW Hamiltonian by 
\be
{\cal L}(\gamma)=-i{\cal U}^\dagger H_{\rm USW}(\mathfrak{u}) {\cal U}
-\gamma\cos^2(2u_0)L  \ ,\quad
\ee
where the unitary transformation ${\cal U}$ is given by \fr{U} and the
parameter $\mathfrak{u}$ is purely imaginary and related to $\gamma$ by
\be
\gamma=-i\frac{\mathfrak{u}}{\cosh\big(2h(u_0)\big)}.
\ee
\subsubsection{Differential equations for correlation functions}
As the jump operators are Hermitian the Lindblad equation implies the
following time evolution for expectation values of (time independent) operators
\bea
\frac{d}{dt}{\rm Tr}\left[\rho(t){\cal O}\right]=-i{\rm
  Tr}\left(\rho(t)[{\cal O},H_0]\right)+\frac{\gamma}{2}
\sum_j {\rm Tr}\left(\rho(t)\ [[L_j,{\cal O}],L_j]\right).
\label{EEQ}
\eea
It is straightforward to verify that the jump operators \fr{LUSW} fulfil
\bea
[L_n,c_j]&=&2\delta_{n,j-1}\sin(u_0)\big(\cos(u_0)c_{j-1}-\sin(u_0)c_j\big)\nn
&+&2\delta_{n,j}\cos(u_0)\big(\cos(u_0)c_{j}-\sin(u_0)c_{j+1}\big).
\eea
This shows that n-particle Green's functions fulfil simple, closed
evolution equations. This is analogous to the case of the imaginary-U
Hubbard model\cite{Medvedyeva}. For example, the single-particle
Green's function
\be
G_{j,k}(t)={\rm Tr}\left[\rho(t) c^\dagger_jc_{k}\right]
\ee
has the following equation of motion
\bea
\frac{d}{dt}G_{j,k}&=&\sum_{\ell,m}K_{j,k}^{\ell,m}G_{\ell,m}\ ,\nn
K_{j,k}^{\ell,m}&=&
\delta_{j,\ell}\delta_{k-1,m}\bigg[i-\frac{\gamma\sin(4u_0)}{2}\bigg]
+\delta_{j,\ell}\delta_{k+1,m}\bigg[i+\frac{\gamma\sin(4u_0)}{2}\bigg]\nn
&-&\delta_{j-1,\ell}\delta_{k,m}\bigg[i-\frac{\gamma\sin(4u_0)}{2}\bigg]
-\delta_{j+1,\ell}\delta_{k,m}\bigg[i+\frac{\gamma\sin(4u_0)}{2}\bigg]\nn
&-&4\gamma\delta_{j,\ell}\delta_{k,m}\cos^2(2u_0)-4\gamma\delta_{j,k}\left[
\sin^2(u_0)M_{j-1}^{\ell,m}-\cos^2(u_0)M_{j}^{\ell,m}\right]\nn
&-&4\gamma\left[\delta_{j-1,k}
\sin(u_0)\cos(u_0)M_{j-1}^{\ell,m}
-\delta_{j,k-1}
\sin(u_0)\cos(u_0)M_{j}^{\ell,m}\right].
\eea
Here we have defined
\be
M_{j}^{\ell,m}=\cos^2(u_0)\delta_{\ell,j}\delta_{m,j}-\sin(u_0)\cos(u_0)\big[
\delta_{\ell,j}\delta_{m,j+1}-\delta_{\ell,j+1}\delta_{m,j}\big]-\sin^2(u_0)\delta_{\ell,j+1}\delta_{m,j+1}.
\ee
\subsection{Maassarani models}
\label{sec:Maa}
In \cite{Maa,Maa2} Maassarani introduced a class of integrable 2n-state
models that generalize the Hubbard model along the lines set out in 
section \ref{sec:intHub} above. We now discuss these models in more
detail. A basis of the local Hilbert space is given by the tensor product 
\be
|a\rangle\otimes|\tilde{a}\rangle\ ,\quad
a,\tilde{a}=1,\dots, n,
\ee
where all states are bosonic,
i.e. $\epsilon_a=0=\epsilon_{\tilde{a}}$. While these models a priori
are generalized spin models they can be related to interacting fermion
models by Jordan-Wigner transformations as is done for a simple case
below. A basis of operators acting on these states is then given by
$e_j^{ab}\tilde{e}_j^{\tilde{a}\tilde{b}}$. In terms of these
(bosonic) operators Maassarani's Hamiltonian reads
\be
H_{\rm Ma,n}(U)=\sum_{j=1}^L
P^{(n)}_{j,j+1}+\widetilde{P}^{(n)}_{j,j+1}+U\big(C_j\widetilde{C}_j-1\big),
\label{HMaa}
\ee
where 
\bea
P^{(n)}_{j,j+1}&=&\sum_{a\in A}\sum_{b\in B}x_{ab}e^{ba}_je^{ab}_{j+1}
+x^{-1}_{ab}e^{ab}_je^{ba}_{j+1}\ ,\nn
C_j&=&\sum_{a\in A}e_j^{aa}-\sum_{b\in B}e_j^{bb}\ .
\label{HMaa2}
\eea
Here the two sets $A$ and $B$ form an arbitrary partition of
$\{1,\dots,n\}$ and $x_{ab}$ are arbitrary complex parameters. In the
following we will simply set them equal to $1$. The
operators $\widetilde{P}^{(n)}_{j,j+1}$ and $\widetilde{C}_j$ are of the
same forms as ${P}^{(n)}_{j,j+1}$ and ${C}_j$ respectively but with
the replacement $e^{ab}_j\rightarrow \widetilde{e}^{ab}_j$. 

Maassarani's models are related to Lindblad equations with Hamiltonians
\be
H_0^{(n)}=-\sum_j\left[\sum_{a\in A}\sum_{b\in B}e^{ba}_je^{ab}_{j+1}+e^{ab}_je^{ba}_{j+1}\right] ,
\ee
and jump operators
\be
L_j=c-C_j.
\label{Maajump}
\ee
In the superoperator formalism the corresponding Liouvillian is
\bea
{\cal L}_{\rm Ma,n}(\gamma)&=&-i(H^{(n)}_0-\widetilde{H}^{(n)}_0)
+\gamma\sum_j\left[C_j\widetilde{C}_j-1\right],
\eea
where $\widetilde{H}^{(n)}_0$ is of the same form as ${H}^{(n)}_0$ but
with $e_j^{ab}$ replaced by $\widetilde{e}_j^{ab}$. This is related to
Maassarani's Hamiltonian by 
\be
{\cal L}_{\rm Ma,n}(\gamma)=i{\cal U}H_{\rm Ma,n}(-i\gamma){\cal
  U}^\dagger\ ,\quad
{\cal U}=\prod_{j=1}^{L/2}\widetilde{C}_{2j}\ .
\ee
\subsubsection{3-state Maassarani model}
The simplest Maassarani model is obtained by considering a local
Hilbert space of three bosonic states. Choosing a decomposition
$A=\{1\}$, $B=\{2,3\}$ gives
\be
H^{(3)}_0=-\sum_{j} e_j^{21}e_{j+1}^{12}+e_j^{31}e_{j+1}^{13}+{\rm
  h.c.}\ .
\ee
In order to fermionize this model we embed it into an enlarged Hilbert
space with four states per site, and then employ the results of
section \ref{ssec:decomp}. This gives
\bea
e_j^{12}=\mathfrak{e}_j^{12}\ \widetilde{\mathfrak{e}}^{11}_j\ ,\qquad
e_j^{13}=\mathfrak{e}_j^{11}\widetilde{\mathfrak{e}}^{12}_j\ .
\eea
Finally we carry out a Jordan-Wigner transformation
\bea
\mathfrak{e}_j^{21}&=&\prod_{\ell=1}^{j-1}(1-2n_{\ell,\up})c^\dagger_{j,\up}\ ,\qquad
\tilde{\mathfrak{e}}_j^{21}=\prod_{\ell=1}^{L}(1-2n_{\ell,\up})\prod_{\ell=1}^{j-1}(1-2n_{\ell,\down})c^\dagger_{j,\down}\ .
\eea
After these transformations the Hamiltonian $H^{(3)}_0$ can be written
in the form
\bea
H^{(3)}_0=-\sum_{j,\sigma} \left[c^\dagger_{j+1,\sigma}c_{j,\sigma}+{\rm
    h.c.}\right]
(1-n_{j,\bar\sigma})(1-n_{j+1,\bar\sigma})
=-{\cal P}\sum_{j,\sigma} \left[c^\dagger_{j+1,\sigma}c_{j,\sigma}+{\rm
    h.c.}\right]{\cal P},
\label{t0}
\eea
where 
\be
{\cal P}=\prod_{j=1}^L(1-n_{j,\up}n_{j,\down})
\label{projop}
\ee
is a projection operator that ensures that all sites are at most
singly occupied. The Hamiltonian \fr{t0} can be viewed as the
$U\to\infty$ limit of the Hubbard model and is sometimes referred to
as the $t-0$ model. In terms of the fermionic operators the jump
operator takes the form  
\be
L_j=1-2(1-n_{j,\up})(1-n_{j,\down})+c.
\ee
Choosing $c=1$ we have
\be
L_j|0\rangle=0\ ,\quad
L_jc^\dagger_{j,\sigma}|0\rangle=2c^\dagger_{j,\sigma}|0\rangle,
\ee
which shows that the bath acts on the charge degrees of freedom. The
Hamiltonian part $H_0^{(3)}$ has a free fermionic spectrum
\cite{t0,Izergin}, but the creation operators of the non-interacting
fermion degrees of freedom are related to the $c_{j,\sigma}^\dagger$
in a non-local way \cite{Kumar,Nocera}. As a result the
single-particle Green's function does not obey a simple evolution
equation. The time evolution is again given by the general expression
\fr{EEQ}, where the relevant commutators are
\bea
[[L_n,c_{j,\sigma}],L_n]{\cal P}&=&-4c_{j,\sigma}\delta_{j,n}\ {\cal
  P}\ ,\nn
{\cal P}\lbrack c_{j,\sigma},H_0^{(3)}\rbrack{\cal P}&=&{\cal P}\big[-(c_{j+1,\sigma}+c_{j-1,\sigma})
-c^\dagger_{j,\bar\sigma}c_{j,\sigma}(c_{j+1,\bar\sigma}+c_{j-1,\bar\sigma})\big]{\cal P}.
\eea
\subsubsection{4-state Maassarani model}
In the 4-state case we can express the $e_j^{ab}$ in terms of two species of Pauli
operators, \emph{cf.} \ref{ssec:decomp}. Choosing $A=\{1,2,3\}$ and
$B=\{4\}$ we then can interpret $H_0^{(4)}$ as the Hamiltonian of a
two-leg spin ladder model
\bea
H_0^{(4)}&=&\sum_{j=1}^L\bigg[\sigma^+_{j}\sigma^-_{j+1}\tau^+_{j}\tau^-_{j+1}
+\sigma^-_{j}\sigma^+_{j+1}\tau^-_{j}\tau^+_{j+1}
+\frac{1}{4}(\sigma^+_{j}\sigma^-_{j+1}+\sigma^-_{j}\sigma^+_{j+1})(1-\tau^z_{j})(1-\tau^z_{j+1})\nn
&&\qquad+\frac{1}{4} (\tau^+_{j}\tau^-_{j+1}+\tau^-_{j}\tau^+_{j+1})(1-\sigma^z_{j})(1-\sigma^z_{j+1})\bigg],
\eea
The jump operators become (setting again $c=1$ in \fr{Maajump})
\be
L_j=2(1-\sigma^z_{j})(1-\tau^z_{j})\ .
\ee

\subsubsection{Bethe Ansatz solution}
The Maassarani models have been solved by Bethe Ansatz in
Ref. \cite{Fomin}. Without loss of generality we restrict our discussion to the
case where the sets $A$ and $B$ in \fr{HMaa2} are given by
\be
A=\{1,2,\dots,p\}\ ,\qquad
B=\{p+1,p+2,\dots,n\}\ .
\ee
The exact eigenstates of $H_{\rm Ma,n}(U)$ are then labelled by good
quantum numbers as follows. The operators 
\be
Q^a=\sum_{j=1}^L e_j^{aa}\ ,\qquad
\tilde{Q}^a=\sum_{j=1}^L \tilde{e}_j^{aa}\ ,\qquad a=1,\dots,n
\ee
commute with $H_{\rm Ma,n}(U)$ and with one another. Hence their
eigenvalues $N_a$, $\tilde{N}_a$ can be used as good quantum numbers.
Following Ref.~\cite{Fomin} we introduce integers
\be
N_A=\sum_{a=2}^p N_a\ ,\quad N_B=\sum_{a=p+1}^{n} N_a\ ,\quad
\tilde{N}_A=\sum_{a=2}^p \tilde{N}_a\ ,\quad
\tilde{N}_B=\sum_{a=p+1}^{n-1} \tilde{N}_a\ ,
\ee
and $N\geq N_A+N_B+\tilde{N}_A+\tilde{N}_B$. We then define sets
\bea
{\cal M}_A&=&\{1,\dots,N_A\}\ ,\quad
{\cal M}_B=\{N_A+1,\dots,N_A+N_B\}\ ,\nn
\tilde{\cal M}_A&=&\{N_A+N_B+1,\dots,N_A+N_B+\tilde{N}_A\}\ ,\nn
\tilde{\cal M}_B&=&\{N_A+N_B+\tilde{N}_A+1,\dots,N_A+N_B+\tilde{N}_A+\tilde{N}_B\}\ ,
\eea
and finally introduce two non-intersecting ordered sets of integers
$1\leq a_j\leq N\leq L$
\be
\mathbb{A}_A=\{a_j|j\in{\cal M}_A\}\ ,\quad
\tilde{\mathbb{A}}_A=\{a_j|j\in\tilde{{\cal M}}_A\}\ ,\quad
\mathbb{A}_A\cap\tilde{\mathbb{A}}_A=\varnothing\ ,\quad
\mathbb{A}_A\cup\mathbb{\tilde{A}}_A\equiv\mathbb{A}.
\ee
By ordered we mean that $a_j<a_{j+1}$ if $a_j,a_{j+1}\in {\mathbb A}_A$ and
similarly for $\tilde{\mathbb{A}}_A$.
The eigenstates of the Liouvillian ${\cal L}_{\rm Ma,n}(\gamma)$ are
then given in terms of rapidities $\{k_1 , \dots , k_{N} \}$,
$ \{ \Lambda_j|j\in{\cal M}_B\}$,
$\{b_m|m\in\tilde{\cal M}_B\}$ and integers
$\{n_1,\dots,n_{\tilde{N}_B-\tilde{N}_{n-1}} \}$,
$\{\bar{n}_1,\dots,\bar{n}_{{N}_B-N_n} \}$ subject to the following
set of Bethe Ansatz equations \cite{Fomin}
\bea
e^{ik_jL}&=&e^{2\pi i\Phi} \prod_{l\in {\cal M}_B}\frac{\Lambda_l - \sin k_j + \gamma}{\Lambda_l - \sin k_j - \gamma}\ ,\quad j\in[1,N]\backslash\mathbb{A} ,\nn
\prod_{\ontop{j=1}{j\notin\mathbb{A}}}^{N}\frac{\Lambda_m - \sin k_j +
  \gamma}{\Lambda_m - \sin k_j - \gamma}&=& e^{2\pi i\Psi}
\prod_{\ontop{l\in{\cal M}_B}{l\neq m}}
  \frac{\Lambda_m - \Lambda_l + 2\gamma}{\Lambda_m - \Lambda_l - 2\gamma}\ ,\quad
  m\in{\cal M}_B\ ,\label{BAEHub}\\
b_\ell^{\tilde{N}_B+\tilde{N}_n}&=&\prod_{j=1}^{\tilde{N}_B-\tilde{N}_{n-1}}e^{2\pi
  i\frac{n_j}{\tilde{N}_B}}\ ,\quad 1\leq
n_1<\dots<n_{\tilde{N}_B-\tilde{N}_{n-1}}\leq \tilde{N}_B\ ,\
\ell\in\tilde{M}_B\ ,\nn  
e^{ik_j(L-N_B)}&=&(-1)^{N_A-1}e^{2\pi i\frac{m_\alpha}{N_A}}\ ,\quad
  m_\alpha\in[1,N_A]\ ,\quad j\in\mathbb{A}\ ,\nn
e^{ik_j(L-\tilde{N}_B-\tilde{N}_n)}&=&(-1)^{\tilde{N}_A-1}e^{2\pi
  i\frac{\tilde{m}_\alpha}{\tilde{N}_A}}\ ,\quad
\tilde{m}_\alpha\in[1,\tilde{N}_A]\ ,\quad j\in\tilde{\mathbb{A}}\ ,
\eea
where we require ${\rm arg}(b_\ell)<{\rm arg}(b_{\ell+1})$ and the
phases $\Phi$ and $\Psi$ are given by
\bea
e^{2\pi i\Phi}&=&(-1)^{\tilde{N}_B+\tilde{N}_n-1}\prod_{m\in
  \tilde{M}_B}b_m\prod_{j\in \tilde{M}_A}e^{-ik_{a_j}}\ ,\nn
e^{2\pi i\Psi}&=&(-1)^{N-N_A-\tilde{N}_A}\prod_{j\in{M}_A}e^{-ik_{a_j}}
\prod_{m\in\tilde{M}_A}e^{ik_{a_m}}\prod_{\ell\in\tilde{M}_B}b_\ell^{-1}
\prod_{s=1}^{N_B-N_n}e^{2\pi i\frac{\bar{n}_s}{N_B}}\ ,\nn
&&1\leq\bar{n}_1<\dots<\bar{n}_{N_B-N_n}<N_B.
\eea
The corresponding eigenvalues of ${\cal L}_{\rm Ma,n}(\gamma)$ are
\be
E=2i\sum_{j\in {\cal M}_B\cup\tilde{\cal M}_B}\cos{k_j}-2\gamma(N_B+\tilde{N}_B+\tilde{N}_n).
\ee
\subsubsection{String solutions and vanishing of the Liouvillian gap
  in the thermodynamic limit}
The first two sets \fr{BAEHub} of the Bethe Ansatz equations are the
same as for the Hubbard model with imaginary interactions strength and
twisted boundary conditions. This ensures that the ``$k$-$\Lambda$ string
solutions'' constructed in \cite{Medvedyeva} are valid solutions for
the n-state Maassarani models as well. A $k$-$\Lambda$ string of
length $m$ corresponds to the following pattern of rapidities
\begin{eqnarray}
k^{(m)}_{\alpha, j} &=& \arcsin(i \Lambda^{(m)}_\alpha - (m - 2j
+2)\gamma'),\nonumber\\
k^{(m)}_{\alpha,j+m} &=& \pi - \arcsin(i \Lambda^{(m)}_\alpha + (m - 2j +2)\gamma') ,
\nonumber\\
\Lambda^{(m)}_{\alpha, j} &=& i\Lambda^{(m)}_\alpha + \gamma (m+1 -
2j)\ ,\qquad 1\le j\le m. \label{T3} 
\end{eqnarray}
Here the string centres $\Lambda^{(m)}_\alpha$ are real and 
$\gamma'=-\gamma\ {\rm sgn}(\Lambda^{(m)}_\alpha)$. 

We now take $N_A=\tilde{N}_A=0$ and consider a Bethe Ansatz state
with a single $k$-$\Lambda$ string of length $m\ll L$. The corresponding
eigenvalue of the Liouvillian is
\be
\epsilon=4{\rm Im}\sqrt{1-(i|\Lambda^{(m)}_\alpha|-m\gamma)^2}-4\gamma m.
\label{epsilon}
\ee
In the framework of
the string hypothesis the equation that fixes the allowed positions of
the string centres $\Lambda^{(m)}_\alpha$ is obtained my ``multiplying
out the string'' \cite{Takahashibook}, which gives
\bea
\exp\left(iL\sum_{j=1}^{2m}k^{(m)}_{j}\right)=e^{2\pi im(2\Phi-\Psi)}.
\eea
Taking logarithms this can be cast in the form
\bea
{\rm sgn}(\Lambda^{(m)})\big[\pi-{\rm arcsin}(i\Lambda^{(m)}+m\gamma)+
{\rm arcsin}(i\Lambda^{(m)}-m\gamma)\big]=\frac{2\pi}{L}
\big(J^{(m)}_\alpha+\varphi\big), 
\label{logBAE}
\eea
where we have defined
\be
\varphi=m(2\Phi-\Psi)\ \text{mod }2\pi.
\ee
For even lattice lengths $L$ the $J^{(m)}_\alpha$ are integers with range
\be
-\frac{L+1-2m}{2}-\varphi<J^{(m)}<\frac{L+1-2m}{2}-\varphi.
\ee
We now focus on the particular sequence of string states characterized
by integers
\be
J^{(m)}_\alpha=\frac{L}{2}-m-\alpha\ ,\quad\alpha=1,2,\dots\ll L.
\ee
In the limit of large system sizes $L\gg 1$ the corresponding string
centres follow from \fr{logBAE}
\be
\Lambda^{(n)}_\alpha=\frac{m\gamma L}{\pi(m+\alpha-\varphi)}+{\cal O}(1).
\ee
Substituting this into our expression \fr{epsilon} for the eigenvalue of the
Liouvillian gives
\be
\epsilon^{(m)}_\alpha=-\frac{2\pi^2}{m\gamma
  L^2}(m+\alpha-\varphi)^2+{\cal O}(L^{-4}).
\ee
This shows that in the large-$L$ limit we have a band of Liouvillian
eigenstates with eigenvalues that scale as $L^{-2}$. This establishes
that the Liouvillian gap vanishes in the thermodynamic
limit. Moreover, the scaling with system size suggests that the
corresponding eigenmodes are diffusive. 
\subsection{\texorpdfstring{$GL(M,N)$ Maassarani models}{Lg}}
\label{sec:GMaa}
As we already mentioned above in section \ref{sec:intHub} the
Shastry-Maassarani construction can be generalized to graded 
magnets based on $GL(M,N)$. Following Ref.~\cite{DFFR} we consider the
class of Hamiltonians
\be
H_{\rm gMa}(U)=\sum_j \Pi^{(n)}_{j,j+1}+\widetilde{\Pi}^{(n)}_{j,j+1}+U\big[C_j\widetilde{C}_j-1\big],
\ee
where
\bea
\Pi^{(n)}_{j,j+1}&=&\sum_{k\neq N,K}\Big[E^{kN}_jE^{Nk}_{j+1}-E^{kK}_jE^{Kk}_{j+1}
+(-1)^{\epsilon_k}(E^{Nk}_jE^{kN}_{j+1}+E^{Kk}_jE^{kK}_{j+1})\Big] ,\nn
C_j&=&1-2E_j^{KK}-2E_j^{NN}\ ,\qquad K=N+M.
\eea
We can relate this to a Lindblad equation with Hamiltonian
\be
H_0=-\sum_j\Pi^{(3)}_{j,j+1},
\ee
and jump operators
\be
L_j=1-C_j.
\ee
\subsubsection{3-state \texorpdfstring{$GL(1,2)$}{Lg} model}
The simplest example is the 3-state model based in $GL(1,2)$. Like in
the case of the 3-state Maassarani model considered above we may
represent the Hamiltonian in terms of canonical spinful fermion
creation and annihilation operators by identifying the three states
per site as 
\be
|1\rangle_j=|0\rangle_j\ ,\quad
|2\rangle_j=c^\dagger_{j,\up}|0\rangle_j\ ,\quad
|3\rangle_j=c^\dagger_{j,\down}|0\rangle_j\ .
\ee
Then $H_0$ can be represented as
\be
H_0=-{\cal P}\sum_{j=1}^L \big(c^\dagger_{j,\up}c_{j+1,\up}-S^+_jS^-_{j+1}+{\rm
  h.c.}\big){\cal P} ,
\ee
where ${\cal P}$ is the projection operator on singly occupied sites
\fr{projop} and $S_j^+=c^\dagger_{i,\up}c_{j,\down}$. This describes
correlated hopping of the up fermions, whereas the down fermions can
only move through spin-flip processes. The jump operator is
\be
L_j=2n_{j,\up}-1\ .
\ee
\section{Other integrable two-leg ladder models}
The generalized Hubbard models considered above are all related to
Lindblad equations with a single jump operator on each bond by virtue
of their integrability structure. There are many other integrable
models that can be represented as two-leg ladders and a question we
have investigated at some length is whether some of them can be
associated with Lindblad equations as well.

\subsection{\texorpdfstring{$GL(N^2)$ magnets}{Lg}}
We now consider generalized spin models on a local Hilbert space with
$N^2$ bosonic states. A well-known class of integrable models is
obtained by taking \cite{Lai,Sutherland}
\bea
H_{GL(N^2)}&=&\sum_{j=1}^L\sum_{\alpha,\beta=1}^{N^2}
E_j^{\alpha\beta}E_{j+1}^{\beta\alpha}\ ,
\eea
where $P_{j,j+1}=\sum_{\alpha,\beta=1}^{N^2}E_j^{\alpha\beta}E_{j+1}^{\beta\alpha}$
is a permutation operator acting on nearest-neighbour lattice sites
\be
P_{j,j+1}|\gamma\rangle_j|\delta\rangle_{j+1}=
|\delta\rangle_j|\gamma\rangle_{j+1}.
\ee
The Hamiltonian $H$ is $GL(N^2)$ symmetric and hence
\be
[H,Q^{\alpha\beta}]=0\ ,\quad
Q^{\alpha,\beta}=\sum_{j=1}^LE_j^{\alpha\beta}\ .
\ee
\subsubsection{Representation as a 2-leg ladder}
The permutation models can be viewed as 2-leg ladders by employing the
decomposition of section \ref{ssec:decomp} for $M=N$. This provides a
representation of the permutation operator as a tensor product
\be
P_{j,j+1}=\left[\sum_{\alpha,\beta=1}^N\et^{\alpha\beta}_j\ \et_{j+1}^{\beta\alpha}\right]
\left[\sum_{\gamma,\delta=1}^Ne^{\gamma\delta}_j\ e_{j+1}^{\delta\gamma}\right].
\label{pjj+1}
\ee
It is clear from the representation \fr{pjj+1} that
\be
[H,{J}^{\alpha\beta}]=0=[H,\widetilde{J}^{\alpha\beta}]=0\ ,
\ee
where
\be
\widetilde{J}^{\alpha\beta}=\sum_{j=1}^L \et_j^{\alpha\beta}\ ,\quad
J^{\alpha\beta}=\sum_{j=1}^L e_j^{\alpha\beta}\ ,\qquad
\alpha,\beta=1,\dots N.
\ee
These operators are related to the $GL(N^2)$ symmetry generators by
\be
J^{\alpha\beta}=\sum_{\gamma=1}^N
Q^{N(\gamma-1)+\alpha,N(\gamma-1)+\beta}\ ,\quad
\widetilde{J}^{\alpha\beta}=\sum_{\gamma=1}^N Q^{N(\alpha-1)+\gamma,N(\beta-1)+\gamma}.
\ee
\subsubsection{Associated Lindblad equation}
Consider now a Lindblad equation with Hamiltonian $H_0$ and two sets
of jump operators $\{L_j\}$ and $\{\ell^{\alpha\beta}_j\}$
\be
H_0=\sum_{\alpha,\beta=1}^N\lambda_{\alpha\beta}J^{\alpha\beta}\ ,\quad
L_j=\left[\sum_{\ab,\bb=1}^Ne^{\ab\bb}_j\ e_{j+1}^{\bb\ab}\right]\ ,\quad
\ell_j^{\alpha\beta}=e_j^{\alpha\beta}\ .
\ee
Noting that $L_j^\dagger L_j=\mathds{1}$ we conclude that the
corresponding Liouvillian is  
\be
{\cal L}=\sum_{\alpha,\beta=1}^{N^2}f_{\alpha\beta}Q^{\alpha,\beta}
+\gamma\sum_{j=1}^L\big(P_{j,j+1}-1\big)\ ,
\label{LGLN}
\ee
where
\be
\sum_{\alpha,\beta=1}^{N^2}f_{\alpha\beta}Q^{\alpha,\beta}=
\sum_{\ab,\bb=1}^N\! -i\lambda_{\ab\bb}[J^{\ab\bb}-\widetilde{J}^{\ab\bb}]
+\gamma_{\ab\bb}
\left[Q^{N(\ab-1)+\ab,N(\bb-1)+\bb}-\frac{J^{\bb\bb}+\widetilde{J}^{\bb\bb}}{2}\right].
\ee
By construction the first term in \fr{LGLN} commutes with the second,
which is $\gamma H_{GL(N^2)}$. As $H_{GL(N^2)}$ is invariant under 
all global $GL(N^2)$ rotations $U$ we conclude that \fr{LGLN} is
integrable for choices of $\gamma_{\alpha\beta}$ and
$\lambda_{\alpha\beta}$ such that
\be
U\sum_{\alpha,\beta=1}^{N^2}f_{\alpha\beta}Q^{\alpha,\beta}U^\dagger=\sum_{\alpha=1}^{N^2}
g_\alpha Q^{\alpha,\alpha}\ ,\quad g_\alpha\in\mathbb{C}.
\ee
\subsubsection{Twisting the boundary conditions}
As we have mention above, in general we need to consider similarity
transformations when trying to ascertain whether a Lindblad equation
is related to an integrable Hamiltonian. A simple example is
provided by considering a Lindblad equation with vanishing Hamiltonian
and jump operators
\be
L_j=\left[\sum_{\ab,\bb=1}^Ne^{i(\varphi_{\bb}-\varphi_{\ab})}
  e^{\ab\bb}_j\ e_{j+1}^{\bb\ab}\right]\ ,\quad\varphi_{\ab}\in\mathbb{R}.
\ee
The corresponding Liouvillian is
\bea
{\cal L}&=&\gamma\sum_{j=1}^L\left[
\sum_{\at,\bt,\ab,\bb=1}^Ne^{i(\varphi_{\bb}-\varphi_{\ab}-\varphi_{\bt}+\varphi_{\at})}
e^{\ab\bb}_j\ e_{j+1}^{\bb\ab}\
\et^{\at\bt}_j\ \et_{j+1}^{\bt\at}-1\right]\nn
&=&\gamma\sum_{j=1}^L\sum_{\alpha,\beta=1}^{N^2}
\left[E_j^{\alpha\beta}E_{j+1}^{\beta\alpha}e^{i(\phi_\beta-\phi_\alpha)}-1\right],
\eea
where we have used the decomposition \ref{ssec:decomp} and fixed the
phases $\phi_\alpha$ by
\be
\phi_\beta-\phi_\alpha=\varphi_{\bb}-\varphi_{\ab}+\varphi_{\at}-\varphi_{\bt}\ ,
\ee
where $\alpha,\beta,\ab,\bb,\at,\bt$ are related by \fr{abat}. To
relate this to the $GL(N^2)$ Hamiltonian we consider the canonical
transformation  
\be
UE_j^{\alpha\beta}U^\dagger={E}_j^{\alpha\beta}e^{-i(\phi_\alpha-\phi_\beta)j}\ ,
\ee
under which the Liouvillian transforms as
\be
U{\cal L}U^\dagger
=\sum_{j=1}^L\sum_{\alpha,\beta=1}^{N^2}\left[
{E}_j^{\alpha\beta}{E}_{j+1}^{\beta\alpha}-1\right],
\ee
where we have imposed twisted boundary conditions
\be
E_{L+1}^{\beta\alpha}=E_1^{\beta\alpha}e^{-i(\phi_\alpha-\phi_\beta)L}\ .
\ee
We conclude that the Liouvillian is related to the integrable $GL(N^2)$
Hamiltonian with twisted boundary conditions
\be
U{\cal L}U^\dagger=\gamma H_{GL(N^2)}\bigg|_{\rm twisted\ bc}.
\ee

\subsubsection{Example: GL(4) spin ladder}
As a specific example let us consider the $GL(4)$ case
\be
H=J\sum_{j=1}^LP_{j,j+1}
+ \frac{h}{2}\left[ Q^{1,1}+Q^{2,3}+Q^{3,2}+Q^{4,4}\right],
\ee
where we have added a particular generalized magnetic field
term. Using \ref{ssec:decomp} we can express this in terms of two
species of Pauli operators, \emph{cf.} \cite{Gritsev} 
\be
H=\frac{1}{4}\sum_{j=1}^LJ\left(\sigma_{j}.\sigma_{j+1}+1\right)\left(\tau_{j}.\tau_{j+1}+1\right) + h\left(\sigma_{j}.\tau_{j}+1\right)\ .
\label{SU4}
\ee
The related Lindblad equation has no Hamiltonian and two sets of jump operators
\bea
L_j=\frac{1}{2}\sum_{a=x,y,z}\sigma^a_j\sigma^a_{j+1}+1\ ,\quad
\{\ell^{(a)}_j=\sigma_j^a|a=x,y,z\}\ .
\eea
The corresponding Liouvillian is
\be
\mathcal{L}=\gamma\sum_{j=1}^L\left(P_{j,j+1}-1\right)
+\gamma'\sum_{j=1}^L(\sigma^x_j\widetilde{\sigma}^x_j-\sigma^y_j\widetilde{\sigma}^y_j+\sigma^z_j\widetilde{\sigma}^z_j-3).
\ee
After a local basis rotation around the y-axis
\be
\tau_j^x=-\widetilde{\sigma}_j^x\ ,\quad
\tau_j^y=\widetilde{\sigma}_j^y\ ,\quad
\tau_j^z=\widetilde{\sigma}_j^z
\ee
this maps onto \fr{SU4} (up to a constant contribution) if we identify
$\gamma=J/4$ and $h=-\gamma'$. 

\subsection{\texorpdfstring{$GL(n_B^2+n_F^2|2n_Bn_F)$ magnets}{Lg}}
\label{sec:graded}
We now turn to particular graded magnets, where we have $n_B^2+n_F^2$
bosonic and $2n_Bn_F$ fermionic states at a given site of the lattice,
where $n_{B,F}\in\mathbb{N}_0$. A much studied family of integrable
models is given by
\cite{Sutherland,KulishSklyanin,Kulish,EKS,EKS1,EK,belliard} 
\be
H=\sum_j\Pi_{j,j+1}+\sum_j\sum_\alpha\lambda_\alpha E_j^{\alpha\alpha},
\ee
where $\Pi_{j,j+1}$ is a graded permutation operator \fr{GPO} and
$\lambda_\alpha$ are generalized chemical potentials. The case
$n_B=n_F=1$ gives the EKS model (a.k.a. supersymmetric extended
Hubbard model). We now employ the decomposition \ref{ssec:decomp} and choose a
tensor product basis for the local Hilbert space as
\be
|\alpha\rangle=|\at\rangle\otimes |\ab\rangle\ ,\quad
\epsilon_\alpha=\epsilon_{\at}+\epsilon_{\ab}\ ,\ \alpha,\ab=1,\dots n_B+n_F,
\ee
where $\alpha=(n_B+n_F)(\at-1)+\ab$. The $E_j^{\alpha\beta}$'s can
then be expressed as
\bea
E^{\alpha\beta}_j=(-1)^{\epsilon_{\bt}(\epsilon_{\ab}+\epsilon_{\bb})}\ \et_j^{\at\bt}\ e_j^{\ab\bb}\ ,
\eea
which in turn leads to the following decomposition of the graded
permutation operator 
\be
\Pi_{j,j+1}=\left[\sum_{\at,\bt=1}^{n_B+n_F}(-1)^{\epsilon_{\bt}}
  \et_j^{\at\bt}e_{j+1}^{\bt\at}\right]
\left[\sum_{\ab,\bb=1}^{n_B+n_F}(-1)^{\epsilon_{\bb}} e_j^{\ab\bb}e_{j+1}^{\bb\ab}\right].
\ee
\subsubsection{Associated Lindblad equation}
Consider now a Lindblad equation with no Hamiltonian and Hermitian
jump operators 
\be
L_j=\sum_{\ab,\bb=1}^N(-1)^{\epsilon_{\bb}}e^{\ab\bb}_j\ e_{j+1}^{\bb\ab}\ ,
\ee
Noting that
\be
L_j^{\dagger} L_j=\mathds{1}\ ,
\ee
we conclude that the corresponding Liouvillian is
\be
{\cal L}=\gamma\sum_{j=1}^L(\Pi_{j,j+1}-1)\ .
\ee
We can slightly generalize this by following the construction for the
$GL(N^2)$ case, e.g. we can add a Hamiltonian
\be
H=\sum_{\ab=1}^{n_B+n_F}\lambda_{\ab}\sum_{j=1}^L e_j^{\ab\ab}\ .
\ee
\subsection{\texorpdfstring{Integrable spin ladder model of Refs \cite{AARS,Albeverio}}{Lg}}
The Hamiltonian of this model can be cast in the form of a two-leg
spin ladder\cite{Gritsev}
\bea
H(J)&=&\frac{1}{4}\sum_{j=1}^L\Big[
\left(\sigma_{j}.\sigma_{j+1}+1\right)\left(\tau_{j}.\tau_{j+1}+1\right) + J\left(\sigma_{j}.\tau_{j}+1\right) \nn
&&\qquad +\left(\sigma_{j}.\tau_{j}+1\right)\left(\sigma_{j+1}.\tau_{j+1}+1\right)
-\left(\sigma_{j}.\tau_{j+1}+1\right)\left(\tau_{j}.\sigma_{j+1}+1\right)\Big].
\label{HGritsev}
\eea

\subsubsection{Associated Lindblad equation}
The Hamiltonian \fr{HGritsev} is related to a Lindblad equation with
no Hamiltonian part and a set of Hermitian jump operators 
\bea
L_j=\sigma_j\cdot\sigma_{j+1}+1\ ,\quad
A^{(a)}_j=\sum_{b,c}\epsilon_{abc}\sigma^b_j\sigma^c_{j+1}\ ,\quad
B^{(a)}_j=\sigma^a_j+\sigma^a_{j+1}\ ,\quad a=x,y,z.
\eea
After a local basis rotation
\be
\tau_j^a\rightarrow \tau^y_j\tau^a_j\tau^y_j\ ,\quad a=x,y,z
\ee
and setting the $\gamma$ parameters to be equal for all jump operator
terms we arrive at a Liouvillian
\bea
\mathcal{L}=4\gamma H(-4)-12L\gamma.
\eea
\section{A comment on continuum limits}
An interesting question is whether  we can take scaling limits and
arrive at Liouvillians described by integrable QFTs. The answer seems
to be negative. Let us consider a lattice model with Hamiltonian
\be
H_0=-t\sum_{j}c^\dagger_{j}c_{j+1}+c^\dagger_{j+1}c_{j}-\mu\sum_j c^\dagger_jc_j\ ,
\ee
and jump operators
\be
L_j=n_j=c^\dagger_{j}c_{j}\ .
\ee
These give rise to a Liouvillian
\bea
{\cal L}&=&-iH_0+i\widetilde{H}_0+\gamma\sum_j
n_j\tilde{n}_j-\frac{1}{2}(n_j+\tilde{n}_j)\ ,
\eea
where $\tilde{H}_0$ is of the same form as $H_0$ but written in terms
of fermion annihilation and creation operators $\tilde{c}_j$ and
$\tilde{c}^\dagger_j$. The sign difference between $\tilde{H}_0$ and
$H_0$ can be removed by a canonical transformation
\be
\tilde{c}_j\rightarrow\tilde{c}_j(-1)^j.
\ee
In analogy of what we do for unitary time evolution we now consider
the scaling limit
\be
t\to\infty\ ,\quad a_0\to 0,\quad
ta_0^2 
\text{ fixed}.
\ee
In this limit lattice fermion operators are replaced by continuum
fields
\be
c_{j}\simeq\sqrt{a_0}\Psi_\up(x)\ ,\quad
\tilde{c}_{j}\simeq\sqrt{a_0}\Psi_\down(x)\ ,\quad x=ja_0.
\ee
The Liouvillian becomes
\bea
{\cal L}&=&\left[i(2t+\mu)-\frac{\gamma}{2}\right]
\int dx\sum_\sigma\Psi^\dagger_\sigma(x)\Psi_\sigma(x)\nn
&+&ita_0^2\int dx\sum_\sigma
\Psi^\dagger_\sigma(x)\partial^2_x\Psi_\sigma(x)
+{\gamma a_0}\int dx \Psi^\dagger_\up(x)\Psi_\up(x)
\Psi^\dagger_\down(x)\Psi_\down(x).
\eea
A problem now occurs in the first term. If $\gamma$ were purely
imaginary we could tune the chemical potential in such a way to ensure
that the prefactor remains finite in the scaling limit. But given that
$\gamma$ is real and positive we cannot take $\gamma\to\infty$, but
must keep it finite in order to have describe states with finite real
parts of their ``energies''. This means that the only scaling limit is
trivial as the interaction term disappears. This is to be a general
feature independent of integrability. 

\section{Some unsuccessful maps}
Most of the integrable ladder models we have considered cannot be
associated in a straightforward way with Lindblad equations. In the
following we present some representative examples.

\subsection{Perk-Schultz models}
\label{sec:q}
As an example we consider the $N=4$ Perk-Schultz model \cite{PerkSchultz,Schultz}
\bea
H_{\rm PS}=J\sum_j\Big[\cosh(\eta)\sum_\al
  E_j^{\al\al}E_{j+1}^{\al\al}+\sum_{\al\neq\beta}
  E_j^{\beta\al}E_{j+1}^{\al\beta}
  +\sgn(\al-\beta)\sinh\eta\ E_j^{\al\al}E_{j+1}^{\beta\beta}\Big].
\eea
This can be viewed as a q-deformation of the GL(4) Hamiltonian
considered above. Using the decomposition \ref{ssec:decomp} we can
rewrite $H_{\rm PS}$ as
\bea
H_{\rm PS}&=&J\sum_jP_{j,j+1}+\frac{\cosh(\eta)-1}{4}
\big(1+\sigma^z_j\sigma^z_{j+1}\big)\big(1+\tau^z_j\tau^z_{j+1}\big)\nn
&&+\frac{J\sinh(\eta)}{4}\sum_j\big(\sigma^z_{j+1}-\sigma^z_j\big)
\big(1+\tau^z_j\tau^z_{j+1}\big). 
\eea
As the spectra of $\sigma^z_{j+1}-\sigma^z_j$ and
$1+\tau^z_j\tau^z_{j+1}$ are different the term in the second line
cannot be related to a jump operator structure in this representation.
\subsection{Higher conservation laws}
\label{sec:HCL}
A well-known way of obtaining integrable spin-ladder models is by
considering higher conservation laws \cite{vladbook,Takahashi}. In
case of the spin-1/2 Heisenberg XXX chain higher conservation laws
$H^{(k+1)}$ can be obtained from the transfer matrix by taking 
logarithmic derivatives at the ``shift point''. By construction we
have $[H^{(k)},H^{(l)}]=0$. The Hamiltonian we want to consider here
is $H(b)=H^{(2)}+bH^{(4)}+{\rm const}$
\cite{Takahashi,zvyagin,klumper}, which takes the form 
\bea
H(b)&=&4\sum_{j=1}^L\Big[ (1-b){\bf S}_j\cdot{\bf S}_{j+1}
+\frac{b}{2}{\bf S}_j\cdot{\bf S}_{j+2}
+2b\left({\bf S}_{j-1}\cdot{\bf S}_{j+1}\right)
\left({\bf S}_{j}\cdot{\bf S}_{j+2}\right)\nn
&&\qquad-2b\left({\bf S}_{j-1}\cdot{\bf S}_{j+2}\right)
\left({\bf S}_{j}\cdot{\bf S}_{j+1}\right)\Big].
\eea
This can be viewed as a zig-zag ladder model by associating all even
(odd) sites with the first (second) leg, which gives 
\bea
H(b)&=&\sum_{j=1}^{L/2}(1-b)
\boldsymbol{\sigma}_j\cdot\left[\boldsymbol{\tau}_{j}+\boldsymbol{\tau}_{j+1}\right]
+\frac{b}{2}\Big\{\boldsymbol{\sigma}_{j}\cdot\boldsymbol{\sigma}_{j+1}\lbrack
\boldsymbol{\tau}_{j}\cdot\boldsymbol{\tau}_{j+1}+\boldsymbol{\tau}_{j+1}\cdot\boldsymbol{\tau}_{j+2}\rbrack\nn
&+&\boldsymbol{\sigma}_j\cdot\boldsymbol{ \sigma}_{j+1}
+\boldsymbol{ \tau}_j\cdot\boldsymbol{ \tau}_{j+1}
-  \boldsymbol{\tau}_{j}\cdot\boldsymbol{\sigma}_{j+1}\
\boldsymbol{\sigma}_{j}\cdot\boldsymbol{\tau}_{j+1}
-\boldsymbol{\sigma}_{j}\cdot\boldsymbol{\tau}_{j+2}\
\boldsymbol{\sigma}_{j+1}\cdot\boldsymbol{\tau}_{j+1}\Big\}.
\eea
This is asymmetric under leg exchange in a way that precludes a direct
relation with a Lindblad equation.

\subsection{Alcaraz-Bariev model}
\label{sec:AB}
The Alcaraz-Bariev two-parameter families of integrable models
\cite{AB} come in two classes denoted by $A^\pm$ and $B^\pm$
respectively. The $B^\pm$ family contains the Hubbard model as a
special limit and this is the only case in which we succeeded in
obtaining an interpretation in terms of a Lindblad equation. We now
discuss why such a relation does not seem to exist in general for
the $A^\pm$ family of models. 
The Hamiltonian of the $A^\pm$ family can be cast in the form
\be
H_A^{(\epsilon)}=\sum_jT_{j,j+1}
+T^{(1)}_{j,j+1}+T^{(2)}_{j,j+1}
+ gT^{(3)}_{j,j+1}
+\cos\theta[S_{j,j+1}-\epsilon T^{(p)}_{j,j+1}+V_{j,j+1}-\epsilon
  U_{j,j+1}]\ ,
\ee
where $g=(1+\epsilon)(1-\sin\theta)$ and
\bea
T_{j,j+1}&=&-e_j^{21}e_{j+1}^{12}+\et_j^{21}\et_{j+1}^{12}+{\rm
  h.c.},\nn
T^{(1)}_{j,j+1}&=&-(e_j^{21}e_{j+1}^{12}-e_j^{12}e_{j+1}^{21})
\et_j^{22}(\epsilon\sin\theta-1)
+(\et_j^{21}\et_{j+1}^{12}-\et_j^{12}\et_{j+1}^{21})
e_j^{22}(\sin\theta-1),\nn
T^{(2)}_{j,j+1}&=&-(e_j^{21}e_{j+1}^{12}-e_j^{12}e_{j+1}^{21})
\et_{j+1}^{22}(\sin\theta-1)
+(\et_j^{21}\et_{j+1}^{12}-\et_j^{12}\et_{j+1}^{21})
e_{j+1}^{22}(\epsilon\sin\theta-1),\nn
T^{(3)}_{j,j+1}&=&-e_j^{21}e_{j+1}^{12}\et_j^{22}\et_{j+1}^{22}+\et_j^{21}\et_{j+1}^{12}e_j^{22}
e_{j+1}^{22}+{\rm h.c.},\nn
S_{j,j+1}&=&e_j^{21}e_{j+1}^{12}\et_j^{12}\et_{j+1}^{21}
+{\rm  h.c.},\nn
T^{(p)}_{j,j+1}&=&-e_j^{21}e_{j+1}^{12}\et_j^{21}\et_{j+1}^{12}+{\rm
  h.c.}\ ,\nn
V_{j,j+1}&=&e^{-2\eta}e_j^{22}\et_{j+1}^{22}+e^{2\eta}\et_{j}^{22}e_{j+1}^{22},\nn
U_{j,j+1}&=&e_j^{22}\et_j^{22}+e_{j+1}^{22}\et_{j+1}^{22}.
\eea
Here we have carried out a unitary transformation
\be
Ue_{j}^{ab}U^\dagger= e_j^{ab} (-1)^{j(a-b)}
\label{hop}
\ee
on the Hamiltonian given in \cite{AB} in anticipation of relating it
to a Liouvillian on a Lindblad equation. We start by noting that we
require $g=0$ for such an interpretation to be possible. The reason is
that the only way to generate $T_{j,j+1}^{(3)}$ is as a ``cross-term''
in $\ell_j \overline{\ell^\dagger_j}$ with
\be
\ell_j=ae_j^{21}e_{j+1}^{12}+b e_j^{12}e_{j+1}^{21}+ce_j^{22}e_{j+1}^{22}.
\ee
However, such jump operators would also generate an unwanted contribution
\be
|c|^2e_j^{22}e_{j+1}^{22}\widetilde{e}_j^{22}\widetilde{e}_{j+1}^{22}.
\ee
As this cannot be cancelled by introducing additional jump operators
and does not feature in $H_A^{(\epsilon)}$ we conclude that we must
have $g=0$.
Next we turn to the cubic terms $T^{(1)}_{j,j+1}$. These must arise
from jump operators of the form
\be
L_j=ae_j^{21}e_{j+1}^{12}+b e_j^{12}e_{j+1}^{21}+ce_j^{22}.
\label{jumpApm}
\ee
These jump operators give rise to inter-species interactions 
\bea
L_j\overline{L}^\dagger_j&=&|a|^2e_j^{21}e_{j+1}^{12}\et_j^{21}\et_{j+1}^{12}
+ab^*e_j^{21}e_{j+1}^{12}\et_j^{12}\et_{j+1}^{21}
+a^*be_j^{12}e_{j+1}^{21}\et_j^{21}\et_{j+1}^{12}\nn
&+&|b|^2e_j^{12}e_{j+1}^{21}\et_j^{12}\et_{j+1}^{21}
+|c|^2e_j^{22}\et_j^{22}+
c^*(ae_j^{21}e_{j+1}^{12}+be_j^{12}e_{j+1}^{21})\et_j^{22}\nn
&+&ce_j^{22}(a^*\et_j^{21}\et_{j+1}^{12}+b^*\et_j^{12}\et_{j+1}^{21}),
\label{LLt}
\eea
and intra-species interactions
\bea
L^\dagger_jL_j
\!&=&\!|a|^2(1-e_j^{22})e_{j+1}^{22}+|b|^2e_j^{22}(1-e_{j+1}^{22})
+|c|^2e_j^{22}-a^*ce_j^{12}e_{j+1}^{21}+c^*ae_j^{21}e_{j+1}^{12},\nn
\overline{L^\dagger_jL_j}\!&=&\!
|a|^2(1-\et_j^{22})\et_{j+1}^{22}+|b|^2\et_j^{22}(1-\et_{j+1}^{22})
+|c|^2\et_j^{22}-ac^*\et_j^{12}\et_{j+1}^{21}+ca^*\et_j^{21}\et_{j+1}^{12}.
\eea
In order to produce the cubic terms in $H^{(\epsilon)}_A$ we require
\be
a=-b\ ,\quad ac^*=1-\epsilon\sin\theta\ ,\quad
ca^*=\sin\theta-1.
\ee
Combining these with the requirement that $g=0$ leads to
\be
\epsilon=\sin\theta=1.
\ee
In this case the $A^{\pm}$ model reduces to free fermions. We
have also investigated whether carrying out a similarity
transformation $SH_{A}^{(\epsilon)}S^{-1}$ with
\be
S=\prod_{j=1}^L \exp\Big(
\varphi e^{22}_j\widetilde{e}^{22}_j+j\left(\varphi_1
e_j^{11}+\varphi_2 e_j^{22}+
\tilde{\varphi}_1 \widetilde{e}_j^{11}+\tilde{\varphi}_2
\widetilde{e}_j^{22}\right)\Big)
\ee
may facilitate a Lindblad interpretation. The answer appears to be negative.
\section{Discussion}
In this work we have reported our findings for a search for
Yang-Baxter integrable Lindblad equations. We have focussed on
translationally invariant situations where jump operators act on bonds
or sites of a one dimensional chain. We have derived a superoperator
representation for lattice models with both fermionic and bosonic
degrees of freedom, and jump operators which can be bosonic or
fermionic. In this representation the Lindblad equation takes the form
of a imaginary time Schr\"odinger equation with a non-Hermitian
``Hamiltonian'' with local density, which can be thought of in terms
of a two-leg ladder model of interacting spins or fermions. We have
then investigated which Yang-Baxter integrable two-leg ladder models
can be related to such Lindblad equations in a ``direct'' way. Our
main result is that a wide class of generalized Hubbard models can be
interpreted as Liouvillians of Lindblad equations. We traced this back
to their integrability structure, which is based on gluing together
certain solutions of the Yang-Baxter equation in a particular
way. Some of the corresponding dissipative models are physically
meaningful, an example being the infinite-U Hubbard model subject to
on-site dephasing noise. As the jump operators in this class of models
are Hermitian, the completely mixed state is a steady state in all
cases. Using the Bethe Ansatz solution we have shown for a subclass of
generalized Hubbard models that the Liouvillian gap vanishes like
$L^{-2}$ as the thermodynamic limit is approached. The corresponding
eigenstates correspond to particle-like ``excitations'' with
quadratic dispersions, which suggests that the late-time behaviour in
these models is likely to be diffusive.

We have identified a few Yang-Baxter integrable Lindblad equations
that are not generalized Hubbard models by showing that certain known
integrable Hamiltonians can be cast in the form of Liouvillians
associated with a Lindblad equation. However, in most cases we have
considered such mappings are not possible. As this is often difficult
to see we have presented a non-trivial case of such a failure in the
Alcaraz-Bariev two-parameter family of integrable models.

We stress that in this work we have focussed on a particular
``direct'' relation between Liouvillians of Lindblad equations and
Hamiltonians of Yang-Baxter integrable models. There are known cases
where it is possible to establish such relationships by means of more
complicated (non-local) maps \cite{Shibata}. Moreover, as we pointed
out in section \ref{ssec:genform}, one ought to allow for similarity
transformations that maintain locality of the Hamiltonian density in
integrable models when trying to establish relations with Lindblad
equations. A systematic way of doing this is by considering
invariances of the Yang-Baxter equation, \emph{cf.} Chapter 12.2.5 of
Ref.~\cite{book}. For example, given a solution $R(\lambda,\mu)\in
{\rm End}(\mathds{C}\otimes\mathds{C})$ of the Yang-Baxter equation
other solutions can be obtained as
\be
\big[V(\mu)\otimes V(\lambda)\big]R(\lambda,\mu)
\big[V^{-1}(\lambda)\otimes V^{-1}(\mu)\big],
\ee
where $V(\lambda)$ is an invertible $n\times n$ matrix. This allows
one to introduce additional free parameters in the resulting
Hamiltonian. The latter will generally be non-Hermitian, but this is
not a problem in the present context of Lindblad equations. It would
be interesting to pursue this line of enquiry further and a good
starting point will be the models successfully related to Lindblad
equations in this work.

In this we work we focussed on identifying integrable Lindblad
equations and only briefly explored using methods of quantum
integrability to obtain physical properties. A good starting point for
this is to determine the spectrum of the Liouvillian, which is given
in terms of the solutions of the relevant Bethe Ansatz
equations. It is well understood that the nature solutions to Bethe
Ansatz equations changes quite substantially when a parameter is made
complex, as this results in the ``scattering phases'' acquiring
magnitudes different from unity. In practice this means that the
structure of solutions to the Bethe Ansatz equations, which is usually
encoded in appropriate string hypotheses, must be revisited and
typically becomes more involved. Even in the simplest case of the
Hubbard model the structure of Bethe Ansatz roots for Liouvillian
eigenstates with eigenvalues that have large real parts and non-zero
imaginary parts appears to be non-trivial. We plan to report on this
issue in a future publication. Ultimately one would like to determine
the dynamics of general Green's functions
\be
   {\rm Tr}\left[\rho(t) E_{j_1}^{\alpha_1\beta_1}\dots
     E_{j_n}^{\alpha_n\beta_n}\right]
\ee
for evolution from a given initial density matrix $\rho(0)$. In some
of the cases discussed above this is relatively simple because the
equations of motion for these Green's functions decouple and for
two-point functions can thus either be integrated numerically or determined
from the exact Liouvillian eigenstates in the two-particle sector
\cite{Eisler}. In cases like the 3-state Maassarani model a more
involved analysis is required and it would be interesting to
investigate this case in more detail.

\paragraph{Acknowledgements}
We are grateful to F. G\"ohmann, H. Katsura and T. Prosen for very helpful
discussions. This work was supported by the EPSRC under grant EP/N01930X.

\begin{appendix}

\section{Structure of the Liouvillian for the most general jump
  operator acting on a bond}
\label{f_const}
The most general two site bosonic jump operator with nearest-neighbour interactions is
\begin{equation}
    L_j=\sum_{\alpha\beta}\left(\lambda_{\alpha\beta}E^{\alpha\beta}_j
    + \lambda_{\alpha\beta}^{'}E^{\alpha\beta}_{j+1}\right) +
    \sum_{\alpha\beta\gamma\delta}
    \mu_{\alpha\beta\gamma\delta}E^{\alpha\beta}_jE^{\gamma\delta}_{j+1}\ .
\end{equation}
This gives rise to interaction terms between the two legs of the ladder
\be
L_j\overline{L_j^\dagger}={\cal I}_j^{(2)}+{\cal I}_j^{(3)}+{\cal
  I}_j^{(4)}\ ,
\ee
where ${\cal I}_j^{(n)}$ involves $n$ Hubbard operators
$E^{\alpha\beta}_j$, $\widetilde{E}^{\alpha\beta}_j$. The interaction
along a single rung of the ladder is
\begin{equation}
\begin{split}
\mathcal{I}^{(2)}_j=\sum_{\substack{\alpha_1\beta_1\\\alpha_2\beta_2}}
&\Big(\lambda_{\alpha_1\beta_1}\lambda^{*}_{\alpha_2\beta_2}E^{\alpha_1\beta_1}_j\widetilde{E}^{\alpha_2\beta_2}_j+ \lambda_{\alpha_1\beta_1}\lambda_{\alpha_2\beta_2}^{'*}E^{\alpha_1\beta_1}_j\widetilde{E}^{\alpha_2\beta_2}_{j+1}\\ &+\lambda_{\alpha_1\beta_1}^{'}\lambda^{*}_{\alpha_2\beta_2}E^{\alpha_1\beta_1}_{j+1}\widetilde{E}^{\alpha_2\beta_2}_j    +\lambda_{\alpha_1\beta_1}^{'}\lambda_{\alpha_2\beta_2}^{'*}E^{\alpha_1\beta_1}_{j+1}\widetilde{E}^{\alpha_2\beta_2}_{j+1}\Big)\ ,
\end{split}
\end{equation}
while the three and four point interactions on a given plaquette are
given by
\bea
\mathcal{I}^{(3)}_j&=&\sum_{\ontop{\alpha_1\beta_1\gamma_1\delta_1}{\alpha_2\beta_2}}
    \mu_{\alpha_1\beta_1\gamma_1\delta_1}E^{\alpha_1\beta_1}_jE^{\gamma_1\delta_1}_{j+1}
    \left(\lambda^{*}_{\alpha_2\beta_2}\widetilde{E}^{\alpha_2\beta_2}_j+ \lambda_{\alpha_2\beta_2}^{'*}\widetilde{E}^{\alpha_2\beta_2}_{j+1}\right)\nn
&+&\sum_{\ontop{\alpha_1\beta_1}{\alpha_2\beta_2\gamma_2\delta_2}}
    \mu^{*}_{\alpha_2\beta_2\gamma_2\delta_2}
    \left(\lambda_{\alpha_1\beta_1}E^{\alpha_1\beta_1}_j+\lambda_{\alpha_1\beta_1}^{'}E^{\alpha_1\beta_1}_{j+1}\right)
    \widetilde{E}^{\alpha_2\beta_2}_j\widetilde{E}^{\gamma_2\delta_2}_{j+1}\ ,\nn
\mathcal{I}^{(4)}_j&=&\sum_{\ontop{\alpha_1\beta_1\gamma_1\delta_1}{\alpha_2\beta_2\gamma_2\delta_2}}
    \mu_{\alpha_1\beta_1\gamma_1\delta_1}\mu^{*}_{\alpha_2\beta_2\gamma_2\delta_2}E^{\alpha_1\beta_1}_jE^{\gamma_1\delta_1}_{j+1}\widetilde{E}^{\alpha_2\beta_2}_j\widetilde{E}^{\gamma_2\delta_2}_{j+1}\ .
\eea
There are also interaction terms along the two legs of the ladder
\bea
L^{\dagger}_jL_j=
\sum_{\beta\gamma}\big[
\big(\sum_\alpha\lambda_{\alpha\beta}\lambda^*_{\alpha\gamma}\big)E^{\gamma\beta}_{j}
+\big(\sum_\alpha
\lambda^{'}_{\alpha\beta}\lambda^{'*}_{\alpha\gamma}\big)E^{\gamma\beta}_{j+1}\big]
+\sum_{\alpha\beta\gamma\delta}\big[f_{\alpha\beta\gamma\delta}E^{\alpha\beta}_{j}
  E^{\gamma\delta}_{j+1}+{\rm h.c.}\big]\ ,\nn
\overline{L^{\dagger}_jL_j}=
\sum_{\beta\gamma}\big[
\big(\sum_\alpha\lambda^*_{\alpha\beta}\lambda_{\alpha\gamma}\big)
\widetilde{E}^{\gamma\beta}_{j}
+\big(\sum_\alpha\lambda^{'*}_{\alpha\beta}\lambda^{'}_{\alpha\gamma}\big)
\widetilde{E}^{\gamma\beta}_{j+1}\big]
+\sum_{\alpha\beta\gamma\delta}\big[f_{\beta\alpha\delta\gamma}
\widetilde{E}^{\gamma\delta}_{j+1}  \widetilde{E}^{\alpha\beta}_{j}
+{\rm h.c.}\big],
\eea
where
\be
f_{\alpha\beta\gamma\delta}=\lambda^{*}_{\beta\alpha}\lambda^{'}_{\gamma\delta}
+\sum_{\eta}\big[\lambda^{*}_{\eta\alpha}\mu_{\eta\beta\gamma\delta}
+\lambda^{'*}_{\eta\gamma}\mu_{\alpha\beta\eta\delta}\big]
+\frac{1}{2}\sum_{\eta\nu}
(-1)^{(\epsilon_{\alpha}+\epsilon_{\beta})(\epsilon_{\eta}+\epsilon_{\gamma})}
\mu_{\nu\beta\eta\delta}\mu^{*}_{\nu\alpha\eta\gamma}.
\ee
\end{appendix}


\end{document}